\documentclass{article}
\usepackage{graphicx,bm,color,amsmath,amssymb}
\usepackage{verbatim}
\usepackage[hidelinks]{hyperref}
\usepackage[numbers,sort&compress,square]{natbib}
\usepackage[a4paper, total={7in, 9in}]{geometry}
\begin{document}
\begin{center}
\Large
Enhanced bacterial swimming speeds in macromolecular polymer solutions \\[3mm]
\large
Andreas Z\"ottl* and Julia M.~Yeomans, \\[3mm]
\normalsize
Rudolf Peierls Centre for Theoretical Physics, 1 Keble Road, Oxford, OX1 3NP, UK \\[3mm]
*E-mail: \href{mailto:andreas.zoettl@physics.ox.ac.uk}{andreas.zoettl@physics.ox.ac.uk}

\end{center}
\normalsize
{\textbf{
The locomotion of swimming bacteria in simple Newtonian fluids can successfully be described within the framework of low Reynolds number hydrodynamics \cite{Lauga2016}.  The presence of polymers in biofluids generally increases the viscosity, which is expected to lead to slower swimming for a constant bacterial motor torque. Surprisingly, however, several experiments have shown that bacterial speeds increase in polymeric fluids \cite{Schneider1974,Berg1979,Martinez2014,Patteson2015}, and there is no clear understanding why.  Therefore we perform extensive coarse-grained simulations of a bacterium swimming in explicitly modeled solutions of macromolecular polymers of different lengths and densities.  We observe an increase of up to 60\% in swimming speed with polymer density and demonstrate that this is due to a depletion of polymers in the vicinity of the bacterium leading to an effective slip. However this in itself cannot predict the large increase in swimming velocity: coupling to the chirality of the bacterial flagellum is also necessary.}

Microorganisms typically move through complex biological environments which contain high-molecular weight polymeric material. Prominent examples include the extracellular matrix, mucosal barriers and polymer-aggregated marine snow \cite{Azam2007,McGuckin2011}. Many explanations have been proposed to describe the increase in speed of bacteria in such polymeric fluids, including viscoelastic effects \cite{Patteson2015}, local shear thinning \cite{Martinez2014}, local shear-induced viscosity gradients \cite{Gomez2016},  polymer depletion \cite{Man2015} or modelling the polymers as a gel-forming network \cite{Berg1979,Magariyama2002}  or a porous medium \cite{Leshansky2009}. Experiments do not, however, yet have the resolution to distinguish between the different theories. Therefore there is a vital role for detailed numerical models that will allow us to understand motion through biologically relevant but rheologically complex, fluids. Drawing on ideas from simulations of polymer hydrodynamics  \cite{Praprotnik2008} and of bacterial locomotion in Newtonian fluids (see for example Refs.~\cite{Phan-Tien1987,Watari2010,Vogel2013,Hu2015}) we simulate a bacterium moving in  suspensions of different polymer density  (Figure~1 and Supplementary Movie 1). Hence we reproduce, and explain, the enhanced swimming speed. 

Swimming bacteria such as \textit{Pseudomponas aeruginosa}, \textit{Helicobacter pylori} or \textit{Eschericia coli} rotate helical flagella attached to their cell body to create a thrust force which moves them forwards  \cite{Lauga2016}. Inspired by the biological swimmers we employ a model bacterium consisting of an elongated cell body of length $2b$ and width $2a$ connected  to a stiff helical flagellum of radius $R$  (Fig.~\ref{Fig:1}a).  Our swimmer is driven by applying a constant motor torque $\mathbf{T}$ to the flagellum and an opposing torque $-\mathbf{T}$ to the body (Fig.~\ref{Fig:1}b). This results in the body rotating with angular velocity $\boldsymbol{\Omega}$, and the counter-rotating flagellum with angular velocity $\boldsymbol{\omega}$ (Fig.~\ref{Fig:1}a) which drives the model cell to swim forwards at an average speed $V$.

The fluid consists of a Newtonian background fluid at viscosity $\eta_0$ and temperature $T$ modelled by multiparticle collision dynamics (MPCD, see Methods). This is coupled to an ensemble of coarse-grained polymers that are modelled as $N$ spherical beads of diameter $\sigma$ connected by quasi-rigid springs of rest length $l_0=\sigma$ and bending stiffness $k_b$. We consider solutions of five different types of polymers (Fig.~\ref{Fig:1}c): 
(i) long and stiff ($N = 12$; $k_b = 120k_B T$); 
(ii) long and semiflexible ($N = 12$; $k_b = 12k_B T$); 
(ii) long and freely jointed ($N = 12$; $k_b = 0$); 
(iv) short and flexible ($N = 4$; $k_b = 0$); 
(v) monomers ($N = 1$).
Fig.~\ref{Fig:1}d shows typical bacterium and polymer configurations at volume fractions $\rho=0.05$ (top) and  $\rho=0.2$ (bottom). The  fluidity (inverse viscosity) of the polymer solution normalised by the viscosity of the background fluid, $\eta_0 / \eta$, is plotted in  Fig.~\ref{Fig:2}a. As expected the fluidity decreases with increasing polymer density, mostly strongly for the longest, stiff polymers and least strongly for the monomers. 

In extensive computer simulations of the dynamics of a model bacterium swimming through the different polymeric fluids (Methods and Supplementary Movie 1) we measure the time- and ensemble-averaged swimmer speeds $V$, body rotation rate $\Omega$, and helix rotation rate $\omega$ projected along the instantaneous swimmer direction.  Counterintuitively we observe that helical microswimmers {\em increase} their swimming speed for all the fluids  considered, by a factor of up to 60\% for the highest polymer densities (Fig.~\ref{Fig:2}b). By contrast the  body and helix rotation rates decrease (Fig.~\ref{Fig:2}c,d).

The large increase in speed must originate from the fluid structure: for a simple continuum fluid the swimming speed  would simply scale inversely with the viscosity  (compare Fig.~\ref{Fig:2}a) and therefore would always decrease with the addition of polymers \cite{Lauga2016}. To understand its origin we measure the time- and ensemble-averaged flow fields  around the swimmer which we express in cylindrical co-ordinates $\mathbf{v}(r,\phi,z) = v_r \hat{\mathbf{r}} + v_\phi \hat{\boldsymbol{\phi}} + v_z \hat{\mathbf{z}}$ where  $\hat{\mathbf{z}}$ lies along the symmetry axis of the swimmer. Figures~\ref{Fig:3}a,d and Supplementary Fig.~1 show the flow fields  around a swimmer in the $r$-$z$ plane normalized by the swimming speed $V$ (a) without polymers and (d) in the presence of long semiflexible polymer filaments  at high density. The leading order far-field  flow for bacteria swimming in Newtonian fluids is known to be a force dipole field \cite{Drescher2011}. This does not change significantly  in the presence of polymers even at high polymer densities.

However, the scaled  tangential  flows   around the rotating body, $\bar{v}_\phi(r)=v_\phi(r)/(a \Omega)$ (Fig.~\ref{Fig:3}b,e and Supplementary Fig.~1), and helix, $\bar{v}_\phi(r)=v_\phi(r)/(R \omega)$ (Fig.~\ref{Fig:3}c,f  and Supplementary Fig.~1), are weakened in the presence of polymers. In Fig.~\ref{Fig:3}g and Supplementary Fig.~2 we plot the decay of $\bar{v}_\phi(r)$ around the helix for different polymer densities, comparing this to the case without polymers [$\bar{v}_\phi^0(r)=v_\phi^0(r)/(R\omega_0)$].  The decrease in magnitude with increasing polymer density is apparent, but note that all the velocity fields decay with the same power law ($\sim r^{-2}$ due to the rotlet field at intermediate distances, but $\sim r^{-1}$ close to the helix because of near-field effects \cite{Balin2017}). This structure of the velocity field shows that the hydrodynamics is not screened, as would be the case in porous media, and that the corresponding Brinkman  theory which has been used, for example, to explain  \textit{C.~elegans} worms swimming in dense wet granular medium \cite{Jung2010}, and has been proposed as an explanation of swimming enhancement for helical bacteria \cite{Leshansky2009}, is not appropriate here. We obtain similar results for the flows around the cell body (Supplementary Fig.~1).

It is instructive to plot the distance-dependent ratio  $\bar{v}_\phi(r)/\bar{v}_\phi^0(r)$ (Fig.~\ref{Fig:3}h) comparing  the tangential velocity in polymeric fluids to that in fluids without polymers.  In the immediate vicinity of the swimmer, the ratio is close to unity; however, it decreases rapidly with $r$ and quickly levels off to a density-dependent constant, $I$. The fact that the ratio converges to $I$ reiterates that the far-field scaling of the flow field is independent of the polymer density, with only the magnitude of the fluid velocity decreasing. On the other hand, in the near-field there is a thin layer in which the flow field decays more quickly with $r$ than in a polymer-free solvent. This sheath of less viscous fluid is a depletion effect: the polymers are less likely to lie near the surface of the flagellum than in the bulk as a consequence of the finite macromolecular size \cite{deGennes1987}.  
   
Indeed we can extract an average depletion layer thickness  $\delta=0.35a$ around the body and  the helix which  is comparable to the size  of a monomer (see Methods). Depletion can be well represented as an apparent slip velocity at the swimmer surface, $u_s^H=(1-I)R\omega$ and $u_s^B=(1-I)a\Omega$ where we distinguish the apparent slip experienced by the helix $H$ and the body $B$ (Methods). The slip velocities increase with increasing polymer density and, interestingly, both collapse to a single curve for all fluids considered when plotted versus $\eta/\eta_0$, the  scaled viscosity of the polymeric fluid  (Fig.~\ref{Fig:3}i,j). This can be well explained  by a two-fluid  model \cite{Man2015,Tuinier2006,Fan2007} where the fluid around the swimmer locally has viscosity $\eta_0$ in the depletion region and $\eta$ outside (black curves in Fig.~\ref{Fig:3}i,j; see also Methods).

However, apparent slip is in itself not sufficient to explain the enhanced swimming speed observed in Fig.~\ref{Fig:2}b since the apparent slip at the surface of driven  spherical colloids \cite{Fan2007}, or slender rods (Supplementary Fig.~3) never leads to movement faster than in polymer-free solvent. So the depletion layer alone cannot explain why bacteria swim faster in a macromolecular solution. We shall now argue that the enhancement in swimming speed also relies on the chirality of the bacterial flagellum: 

We use Resistive Force Theory (RFT) and locally approximate a helix segment by a slender rod. In RFT the viscous force per unit length $\mathbf{f}$ opposing the motion of a segment is split into a parallel and a perpendicular component \cite{Gray1955,Chwang1971}, $\mathbf{f} = -\xi_{||}\mathbf{u}_{||} - \xi_{\perp}\mathbf{u}_{\perp}$, with anisotropic friction coefficients $1 < \xi_{\perp} / \xi_{||}<2$ and local segment velocity $ \mathbf{u}= \mathbf{u}_{||} + \mathbf{u}_{\perp}$ (Supplementary Fig.~3). The slip acts tangientially along the flagellum, so to model the depletion we  include a slip velocity $u_s^l$ which reduces the parallel component of the helix velocity,
\begin{equation}
\mathbf{f} = -\xi_{||}(u_{||}-u_s^l)\hat{\mathbf{t}} - \xi_{\perp}\mathbf{u}_{\perp},
\label{Eq:1}
\end{equation}
where $\hat{\mathbf{t}}=\cos\alpha\mathbf{\hat{z}} + \sin\alpha\boldsymbol{\hat{\phi}}$ is the unit vector tangential  to the helix and $\alpha$ is its pitch angle. Integrating  gives the relation between the total force $F$ and torque $T$ acting on the helix and its velocity $V_H$ and  angular velocity $\omega_H$  (see Methods for details):
\begin{eqnarray}
V_H &=& \mu_t F + \mu_{tr} T +  R\omega_H\bar{u}_s / \tan\alpha, \label{Eq:1ab} \\
\omega_H &=& \mu_{tr} F +\mu_{r} T +  \omega_H\bar{u}_s,\label{Eq:1bb}
\end{eqnarray}
where $\bar{u}_s=u_s/(R\omega_H)=u_s^l\sin\alpha/(R\omega_H)$ is the normalized apparent slip obtained from the decay of the azimuthal flow field, see Fig.~\ref{Fig:3}j. The  translational, rotational and translation-rotation-coupling  mobilities, $\mu_t$, $\mu_r$, and $\mu_{tr}$ respectively, are all $>0$ and $\propto \eta^{-1}$. As the swimmer's helix is driven only by a torque we put $F=0$. Eliminating $\omega_H$ from Eqs.~(\ref{Eq:1ab}) and~(\ref{Eq:1bb}) gives an expression for the helix velocity relative to the polymer-free solvent velocity, $V_H^0 = \mu_{tr}^0 T$:
\begin{equation}
\frac{V_H}{V_H^0} = \frac{\eta_0}{\eta}\left( 1 +  \frac{1 }{\tan\alpha} \frac{\bar{u}_s}{1-\bar{u}_s} \frac{R \mu_{r}}{\mu_{tr}}   \right).
\label{Eq:v0}
\end{equation}
The relative swimming speed given by Eq.~(\ref{Eq:v0}) depends on two terms: While the first factor, $\eta_0/ \eta$, decreases with viscosity, the second factor increases with viscosity due to the increase of the scaled slip velocity $\bar{u}_s$ with $\eta$ (Fig.~\ref{Fig:3}j). The competition between these two terms  can lead to a speed enhancement for sufficiently large depletion layer thickness and moderate viscosities, as is the case in our system.  Fig.~\ref{Fig:4}a shows how the change in speed, predicted by Eq.~(\ref{Eq:v0}), varies with the scaled viscosity of the polymer solution, and with the thickness of the depletion layer $\delta$. 

Including the counterrotating body and solving the coupled body-helix model gives only a small quantitative correction to this argument (Methods). This is apparent in Fig.~\ref{Fig:4}b where we  compare the dependence of swimmer speed on viscosity obtained from the simulations to analytic results from the torque driven helix model (Eq.~(\ref{Eq:v0}), black dashed line) and the model including the swimmer body (black curve). For comparison we also show the very different behaviour of the swimmer speed in the absence of any depletion effect (red line). The analytic results, which have no free fit parameters, are in excellent qualitative  agreement with the simulations. An exact quantitative match would be fortuitous because of our use of RFT which is known to be approximate for the helix geometry \cite{Rodenborn2013}.  Note that the viscosity ratio $\eta / \eta_0 \approx 2.7$ leads the greatest enhancement in swimming speed even for larger depletion layer thickness, see Fig.~\ref{Fig:4}a. For the longer polymers shear-induced  stretching of polymers near the helix locally induces shear-thinning, but this and viscoelastic effects are minor contributions to the swimming performance \cite{Liu2009,Spagnolie2013,Gomez2016} compared to the polymer depletion, which depends on the value of viscosity in the  bulk.

In our simulations the depletion layer thickness is approximately $0.35a$, and depends only weakly on the polymer type. Normally in coarse-grained simulations the bead size $\sigma$ is interpreted, following the de Gennes `blob' picture \cite{deGennes1979}, as approximating the polymer radius of gyration. However for $\delta>0.3a$ -- which is the regime we are in here -- the slip velocity depends very weakly on the depletion layer thickness so details of exactly how this is resolved by the model are unimportant. While swimming in a continuum viscous fluid without a polymer depletion layer always reduces the swimming speed, a relatively thin polymer-free layer around the swimmer can reverse this effect.

\section*{Methods}
\subsection*{Bacterium model}
The cell body is modeled by a hard superellipsoid \cite{Barr1981}. Its surface at time $t=0$ is defined by  
\begin{equation}
\left[\left(\frac{x-x_0}{a}\right)^{2/ \epsilon_2} + \left(\frac{y-y_0}{a}\right)^{2/ \epsilon_2}      \right]^{\epsilon_2/ \epsilon_1} + \left(\frac{z-z_0}{b}\right)^{2/ \epsilon_1} =1
\end{equation}
with $\epsilon_1=0.5$, $\epsilon_2=1$, $a=2\sigma$, $b=4\sigma$ and an initial body position $x_0=0$, $y_0=0$, $z_0=15.5\sigma$. The right-handed helical tail consists of 27   beads of diameter $\sigma$ placed at positions $\mathbf{R}_i$,
\begin{equation}
\mathbf{R}_i(s_i) = \left( R\left[1-e^{-(s_i/l_s)^2} \right]\cos s_i, R\left[1-e^{-(s_i/l_s)^2} \right]\sin s_i, p s_i + z_0 + b + \sigma/2 \right),
\end{equation}
with pitch parameter $p=\sigma$, radius $R=2\sigma$, Hidgeon length  \cite{Higdon1979} $l_s=3\sigma$, $s_1=0$ and the other $s_i$ such that the helix beads are just touching,  see Fig.~\ref{Fig:1}a. The total helix length is $L\approx 14\sigma$, and the average pitch angle is $\alpha =1.0$ defined by $\tan\alpha=R(1-e^{-(s/l_s)^2})/p$. The helix is connected to 3 auxiliary basis beads (see also \cite{Reigh2012}) at positions $(\sigma/ \sqrt{2},0,-\sigma/ \sqrt{2})$, $(0,-\sigma/ \sqrt{2},-\sigma/ \sqrt{2},)$ and  $(-\sigma/ \sqrt{2},0,-\sigma/ \sqrt{2})$. The motor torque $\mathbf{T}$ is implemented by applying forces $\mathbf{F}=200(k_BT/\sigma) \mathbf{\hat{l}} \times \mathbf{\hat{z}}$ and  $-\mathbf{F}$ to the 1st and 3rd auxilliary bead, respectively, where $ \mathbf{\hat{l}}=\mathbf{l}/l$ is their normalized separation vector  such that the total torque on the helix is $\mathbf{T}=\mathbf{l} \times \mathbf{F}$. In order to simulate a torque-free swimmer the torque $-\mathbf{T}$ is transferred to the cell body. All beads are connected by quasi-rigid springs  modeled by a bond potential, 
\begin{equation}
V_\text{bond}^H = \frac 1 2 k_\text{bond}\sum_{i=2}^{30}(|\Delta \mathbf{R}_i|-l_0)^2
\label{Eq:VHBond}
\end{equation}
with  $\Delta \mathbf{R}_i = \mathbf{R}_i - \mathbf{R}_{i-1}$, rest length $l_0=\sigma$, and  bond potential strength $k_\text{bond}=10^5k_BT$. The helix is almost stiff but we allow some flexibility (compare Ref.~\cite{Reigh2012}) by using a bending potential 
\begin{equation}
V_\text{bend}^H = \frac 1 2 k_{\text{b}}\sum_{i=3}^{30}\left(\frac{\Delta \mathbf{R}_i \cdot \Delta \mathbf{R}_{i-1}}{|\Delta \mathbf{R}_i| |\Delta \mathbf{R}_{i-1}|} -\cos\theta_i\right)^2
\end{equation}
where $k_b=2\times 10^5k_BT$ and $\theta_i$ are the initial bending angles between  three consecutive beads \cite{Allen1989}, and a torsion potential,
\begin{equation}
V_{\text{tors}}^H = \frac 1 2 k_{\text{tors}} \sum_{i=4}^{30} \left(\frac{(\Delta \mathbf{R}_i \times \Delta \mathbf{R}_{i-1} )
 \cdot (\Delta \mathbf{R}_{i-1} \times \Delta \mathbf{R}_{i-2})}{|\Delta \mathbf{R}_i \times \Delta \mathbf{R}_{i-1} |
 |\Delta \mathbf{R}_{i-1} \times \Delta \mathbf{R}_{i-2}|}  - \cos\phi_{i}\right)^2
\end{equation}
where $k_{\text{tors}}=10^5k_BT$ and $\phi_i$ are the initial torsion angles between  four consecutive beads \cite{Allen1989}. The axis of the helix is kept parallel to the body orientation by using additional harmonic potentials.

\subsection*{Polymer model}
Each polymer is modeled by $N$ beads (monomers) of diameter $\sigma$ located at positions $\mathbf{r}_i$, $i=1,\dots,N$ which are again connected by a bond potential
\begin{equation}
V_\text{bond}^P = \frac 1 2 k_\text{bond}\sum_{i=2}^{N}(|\Delta \mathbf{r}_i|-l_0)^2
\end{equation}
with  $\Delta \mathbf{r}_i = \mathbf{r}_i - \mathbf{r}_{i-1}$ and the same $l_0$ and $k_\text{bond}$ as used for the helix. A bending potential
\begin{equation}
V_\text{bend}^P = \frac 1 2 k_\text{b}\sum_{i=3}^{N}\left(\frac{\Delta \mathbf{r}_i \cdot \Delta \mathbf{r}_{i-1}}{|\Delta \mathbf{r}_i| |\Delta \mathbf{r}_{i-1}|} -1\right)^2
\label{Eq:BendP}
\end{equation}
with $k_b=\{0,12k_BT,120k_BT\}$ is included to simulate flexible, semi-flexible and stiff polymers, respectively. A purely repulsive soft WCA potential \cite{Weeks1971} is used between pairs of polymer beads  separated by a distance $r$,
\begin{equation}
 V_{\text{WCA}}(r) = 
\left\{ \begin{array}{rl}
 4\epsilon \left[ {\left( \frac{\sigma^\ast}{r} \right)}^{12} - {\left( \frac{\sigma^\ast}{r}\right)}^{6} \right]  + \epsilon,
  & \textnormal{for } r < 2^{1/6}\sigma^\ast \\
0, & \textnormal{for } r \ge 2^{1/6}\sigma^\ast.  \end{array} \right\}, 
\label{Eq:WCA}
\end{equation}
with $\epsilon=k_BT$ and $\sigma^\ast=\sigma/2^{1/6}$. We use the same potential between polymer and helix beads, and a purely repulsive potential between polymer beads and the cell body. We simulate polymeric fluids at different volume fractions $\rho=\{0.01,0.05,0.1,0.2\}$ with $\rho= N_pN\pi\sigma^3/(6V)$ where $N_p$ is the number of polymers and $V=S_xS_yS_z$ the simulation domain volume with $S_x=66\sigma$, $S_y=66\sigma$ and $S_z=132\sigma$. The polymers are initially randomly distributed  ($x\in \{-S_x/2, S_x/2 \}$, $y\in \{-S_y/2, S_y/2 \}$, $z\in \{-S_z/2, S_z/2 \}$) but are not allowed to overlap with the bacterium.

\subsection*{MPCD-MD simulations}
The Newtonian background fluid is simulated using multiparticle collision dynamics (MPCD). This is a coarse-grained solver of the Navier Stokes equations which naturally includes thermal fluctuations \cite{Kapral2008,Gompper2009}. The fluid is represented by point-like  effective fluid particles of mass $m$, and their dynamics is modeled by alternating streaming and collision steps. In the streaming step fluid particles move ballistically for a time $\delta t$ so that their positions $\mathbf{x}_i$ are updated to
\begin{equation}
\mathbf{x}_i (t + \delta t) = \mathbf{x}_i (t) + \mathbf{v}_i(t) \delta t
\end{equation}
where $\mathbf{v}_i$ are their velocities. They are then sorted into cubic cells of length $h=\sigma$ and, in the collision step, all particles in a cell exchange momentum according to
\begin{equation}
\mathbf{v}_i (t + \delta t) = \mathbf{v}_{\xi} (t) + \mathbf{v}_\text{rand}(t) + \mathbf{v}_P(t) + \mathbf{v}_L(t)
\end{equation}
where $\mathbf{v}_{\xi}$ is the instantaneous average velocity in the cell, $\mathbf{v}_\text{rand}$ is a random velocity drawn from a Maxwell-Boltzmann distribution at temperature $T$, and $\mathbf{v}_P $ and  $\mathbf{v}_L $ ensure local linear and angular momentum conservation, respectively  \cite{Gompper2009}.  We use $\delta t = 0.02 \sqrt{ m h^2 /k_B T}$ and a  fluid particle number density $n = 10h^{-3}$ in order to model viscous flow at  low Reynolds number.

To simulate the molecular dynamics (MD) of the polymers and the bacterium in the MPCD fluid we use a hydrid MPCD-MD scheme \cite{Gompper2009}. Within the streaming step the positions and velocities of the  polymer and helix beads are updated by determing the forces from the potentials [Eqs.~(\ref{Eq:VHBond}) - (\ref{Eq:WCA})] and using a Velocity Verlet algorithm \cite{Allen1989} with time step $\delta t_P = 0.002\sqrt{ m h^2 /k_B T}$ for the polymer beads, $\delta t_H = 0.0002\sqrt{ m h^2 /k_B T}$ for the helix beads and $\delta t_B = 0.02\sqrt{ m h^2 /k_B T}$ for the cell body. Fluid particles interact with the cell body by applying a bounce back rule, and momentum and angular momentum are exchanged accordingly. Polymer and helix beads, which have masses $m_P=m_H=10m$, are coupled to the fluid by including them in the collision step \cite{Malevanets2000}. In order to accurately resolve the flow fields near the cell body we use virtual particles inside the cell body which contribute to the collision step \cite{Gompper2009}. 

In total we simulate the dynamics of  $\approx 5.75\times 10^6$ MPCD fluid particles and of up to $\approx 2.2\times 10^5$ polymer beads (e.g.\ $\approx 18300$ polymers of length $N=12$ at density $\rho=0.2$) for a total simulation time $\Delta t=6000\sqrt{ m h^2 /k_B T}$ corresponding to $3\times 10^5$ streaming and collision steps.  

\subsection*{Shear viscosity measurement}
We use a simple numerical method, following that proposed in Ref.~\cite{Backer2005} for Newtonian fluids, to evaluate the shear viscosity of the polymeric fluids in the absence of the bacterium. The  particles are subjected to a constant acceleration force $\mathbf{f}_a$ in one half of the simulation box and to $-\mathbf{f}_a$ in the other half. Even in the absence of any walls this results in a periodic Poiseuille flow profile  at sufficiently low shear rates. The maximum velocity of the profile, averaged over time and simulation runs, is linearly related to the inverse viscosity \cite{Backer2005}. 

We can estimate the hydrodynamic radius of a monomer, $a_H$, by comparing $\eta(\rho)/ \eta_0$ to the Einstein approximation for the viscosity of a suspension of spheres, $\eta(\rho^\ast)/ \eta_0=1+\frac 5 2 \rho^\ast$ where $\rho^\ast \sim a_H^{-3}$. A linear fit for the viscosity of monomers from our simulations yields $\eta(\rho)/ \eta_0 \approx 1+1.0\rho^\ast$ with $\rho \sim (\sigma/2)^{-3}$. Comparing the two expressions leads to $a_H=0.37\sigma$. 

\subsection*{Two-fluid model for apparent slip velocity}
We use  a simple analytical model to predict the apparent slip velocities near the rotating body and  helix. We find that a two-fluid model around a rotating sphere fits  our data for the slip velocities quantitatively. This model was introduced by Tuinier et al.\ \cite{Tuinier2006,Fan2007} to explain polymer depletion effects around driven colloids.

In the absence of polymers the flow field around the equator of a rotating sphere of radius $a$ with angular velocity $\Omega$ is  simply  $v_\phi^0(r)=a^3\Omega / r^2$. In the presence of polymers we assume that the fluid region can be divided into an inner layer ($a<r<r_s$) with the viscosity of the background fluid  $\eta_0$ and an outer layer ($r>r_s$) with the bulk viscosity of the polymeric fluid $\eta$  \cite{Fan2007} (Supplementary Fig.~4). The flow field then becomes
\begin{equation}
    \begin{cases}
          v^{in}_\phi(r) =  \frac{M a^3 \Omega}{r^2} - N \Omega r     , & \text{for}\ a<r<r_s,\\
        v^{out}_\phi(r) =   \frac{I a^3 \Omega}{r^2}  , & \text{for}\ r>r_s
    \end{cases}   
\label{Eq:uphi}
\end{equation}
with dimensionless constants
\begin{equation}
M = \frac{1}{ \left(   \frac{r_s^3}{a^3} -1 + \frac{\eta_0}{\eta}   \right)  }\frac{r_s^3}{a^3}, \qquad
N = \frac{(1- \frac{\eta_0}{\eta})}{ \left(   \frac{r_s^3}{a^3} -1 + \frac{\eta_0}{\eta}   \right)  }, \qquad
I = \frac{1}{ \left(   \frac{r_s^3}{a^3} -1 + \frac{\eta_0}{\eta}   \right)  } \frac{r_s^3}{a^3}\frac{\eta_0}{\eta}
\label{Eq:TwoF3}
\end{equation}
which only depend on the viscosity ratio $\eta / \eta_0$ and $r_s/a$ where the latter is related to the scaled depletion layer thickness $\delta/a=r_s/a-1$. Note that $v^{out}_\phi / v_\phi^0=I$.

We define the apparent slip velocity as (Supplementary Fig.~4) 
\begin{equation}
u_s^B  = v^{in}_\phi(a)- v^{out}_\phi(a)=(1-I)a\Omega = a\Omega \left(1- \frac{1}{ \left(   \frac{r_s^3}{a^3} -1 + \frac{\eta_0}{\eta}   \right)  } \frac{r_s^3}{a^3}\frac{\eta_0}{\eta}   \right).
\end{equation}
Supplementary Fig.~5 shows that, for $r_s \gtrsim 1.3a$, the depletion layer thickness  has little effect on the apparent slip which approaches a constant, $(1-\eta_0/\eta)a\Omega <a\Omega$, that only depends on $\eta / \eta_0$.  Similarly, the apparent slip near the helix can  be well described  by  $u_s^H=u_s^l\sin\alpha=(1-I)R\omega\sin{\alpha}$ where $\sin\alpha$ is a geometrical factor that accounts for the fact that the helix tangent $\mathbf{\hat{t}}$ is not parallel to $\boldsymbol{\hat{\phi}}$.

\subsection*{Estimation of depletion layer thickness}
A pathology of the two-fluid model is the kink in the velocity profile at $r_s$ (Supplementary Fig.~4). In the simulations $v_\phi(r)$ decays smoothly and therefore a definition is needed for $r_s$ and hence $\delta$ to enable comparison to the model. We choose the distance when $\bar{v}_\phi/ \bar{v}_\phi^0$ decays to halfway between 1 and $I$, namely $ \bar{v}_\phi(a+\delta)/ \bar{v}_\phi^0(a+\delta)=(1+I)/2 $.

\subsection*{Resistive Force Theory (RFT) for the helix including apparent slip}
We use RFT including apparent slip [Eq.~(\ref{Eq:1})] and derive Eqs.~(\ref{Eq:1ab}) and (\ref{Eq:1bb}). We consider a right-handed helical flagellum moving in time $t$ with velocity $V_H$ and rotating with angular velocity $\omega_H$ along the $z$ axis,
\begin{equation}
\mathbf{R}(s,t) = (R \cos(s\sin\alpha/R+\omega_H t),  R \sin(s\sin\alpha/R+\omega_H t), s \cos\alpha + V_H t)
\end{equation}
where  $R$ is the radius and $\alpha$ the pitch angle, $s\in (0,l_c)$ with $l_c$ the contour length. The tangential vector is $\hat{\mathbf{t}}=d\mathbf{R}/ds=\cos\alpha\mathbf{\hat{z}} + \sin\alpha\boldsymbol{\hat{\phi}}$, and the helix segment velocity given by $\mathbf{u} = d\mathbf{R}/dt = R\omega_H\boldsymbol{\hat{\phi}} + V_H\mathbf{\hat{z}}$. Integrating the forces $F=-\hat{\mathbf{z}}\cdot\int\mathbf{f}ds$ and torques $T=-\hat{\mathbf{z}}\cdot \int R\hat{\mathbf{r}}\times \mathbf{f}ds$ along the helix, and solving for $V_H$ and $\omega_H$ results in Eqs.~(\ref{Eq:1ab}) and (\ref{Eq:1bb}) with the conventional helix mobilities \cite{Rodenborn2013}
\begin{equation}
\mu_t = \frac{\xi_{||}+\xi_\perp + (\xi_\perp - \xi_{||})\cos 2\alpha}{2l\xi_{||}\xi_\perp}, \quad\quad
\mu_{tr} = \frac{(\xi_\perp - \xi_{||})\cos \alpha\sin\alpha}{lR\xi_{||}\xi_\perp}, \quad\quad
\mu_r = \frac{\xi_{||}+\xi_\perp - (\xi_\perp - \xi_{||})\cos 2\alpha}{2lR^2\xi_{||}\xi_\perp}.
\end{equation}
We estimate the friction coefficients using the classical formulas from Gray and Hancock \cite{Gray1955},
\begin{equation}
\xi_{||} = \frac{2\pi \eta}{\ln\frac{2\lambda}{a_H}-\frac 1 2}, \quad\quad\quad\quad \xi_{\perp} = \frac{4\pi \eta}{\ln\frac{2\lambda}{a_H}+\frac 1 2},
\end{equation}
where $\lambda=2\pi p $ and  $a_H$ is approximated using the hydrodynamic radius of a helix bead. Thus for our helix geometry we obtain $\xi_{\perp} / \xi_{||} = 1.50$.  Solving Eq.~(\ref{Eq:1bb}) for $\omega_H$ and substituting into Eq.~(\ref{Eq:1ab}) gives 
\begin{equation}
V_H = \left(\mu_t +  \mu_t^s \right) F  + \left(\mu_{tr} + \mu_{tr}^s  \right) T   \label{Eq:3bb}
\end{equation}
where we have introduced the slip-induced effective mobilities
\begin{equation}
\mu_t^s   =  \frac{1 }{\tan\alpha} \frac{\bar{u}_s}{1-\bar{u}_s} R\mu_{tr}, \quad\quad\quad\quad
\mu_{tr}^s = \frac{1 }{\tan\alpha} \frac{\bar{u}_s}{1-\bar{u}_s} R\mu_{r}. 
\label{Eq:56}
\end{equation}
These are positive, as expected, as the slip makes it easier to move for a given force and torque. Putting $F=0$ in Eq.~(\ref{Eq:3bb}),  combining Eqs.~(\ref{Eq:3bb}) and (\ref{Eq:56}), and dividing by the velocity of the helix  in the background fluid gives Eq.~(\ref{Eq:v0}).

\subsection*{Propulsion speed model for bacterium}
We determine the velocity of a model bacterium where a rotating helix is coupled to a rigid body including apparent slip near the helix and near the body. Following  the case without slip \cite{Rodenborn2013,Martinez2014}, we neglect hydrodynamic coupling between body and helix and write the linear relations between propulsion velocity $V$, body and helix angular velocities $\Omega$ and $\omega$, applied constant motor torque $T$ and propulsion force $F$, 
\begin{eqnarray}
\begin{pmatrix}
V \\
\Omega \\
\end{pmatrix}
=
\begin{pmatrix}
\mu_{t,\text{eff}}^{B} & 0 \\
0 &\mu_{r,\text{eff}}^{B} \\
\end{pmatrix}
\cdot
\begin{pmatrix}
F \\
-T \\
\end{pmatrix}
\label{Eq:FullM1}\\
\begin{pmatrix}
V-\bar{u}_sR\omega/ \tan\alpha \\
\omega (1-\bar{u}_s) \\
\end{pmatrix}
=
\begin{pmatrix}
\mu_t & \mu_{tr} \\
\mu_{tr} &\mu_r \\
\end{pmatrix}
\cdot
\begin{pmatrix}
-F \\
T \\
\end{pmatrix}
\label{Eq:FullM2}
\end{eqnarray}
where $\mu_{t,\text{eff}}^{B}=f_t\mu_t^B $ is the effective translational mobility of the head which includes a slip factor $1<f_t< 4/3$ \cite{Fan2007} and we use the  mobility $\mu_t^B $ for a prolate ellipsoid of length $2b$ and width $2a$ to approximate our superellipsoidal shape \cite{KimKarila}. $\mu_{r,\text{eff}}^{B}$ is the effective rotational mobility of the head which does not influence the  velocity of the bacterium.  Solving Eqs.~(\ref{Eq:FullM1}) and (\ref{Eq:FullM2})  and comparing to the polymer-free solvent case ($\bar{u}_s=0$, $\eta = \eta_0$, $f_t=1$) leads to a speed enhancement
\begin{equation}
\frac{V}{V_0} =\frac{\eta_0}{\eta}\left(1 + \frac{\mu_{tr}^s}{\mu_{tr}}\right)   \frac{\mu_{t,\text{eff}}^B+f_t\mu_t}{\mu_{t,\text{eff}}^B + \mu_t + \mu_t^s} = \frac{V_H}{V_H^0} f_B.
\label{Eq:FullM3}
\end{equation}
Note that $V/V_0$ can be written as a product of the speed enhancement for a helix driven by a torque [$V_H / V_H^0$, Eq.~(\ref{Eq:v0})] and a factor $f_B\gtrsim 1$ which depends on both body and helix translational mobilities and is only responsible for a small increase in bacterium speed (see Fig.~\ref{Fig:4}).\\

\section*{Acknowledgements}
We are grateful to Tyler Shendruk for helpful discussions and comments on the manuscript. We acknowledge discussions with Wilson Poon, Andrew Balin and Amin Doostmohammadi. This project has received funding from the European Union's Horizon 2020 research and innovation programme under the Marie Sk\l{}odowska-Curie grant agreement No.\ 653284. The authors would like to acknowledge the use of the University of Oxford Advanced Research Computing (ARC) facility in carrying out this work. 

\section*{Author contributions}
A.Z.\ and J.M.Y.\ designed the research. A.Z.\ developed the MPCD simulation code, performed the simulations, analyzed the data and developed the theory. A.Z.\ and J.M.Y.\ developed  the results and wrote the paper.


\begin{figure}[bt]
\begin{center}
\includegraphics[width=.8\columnwidth]{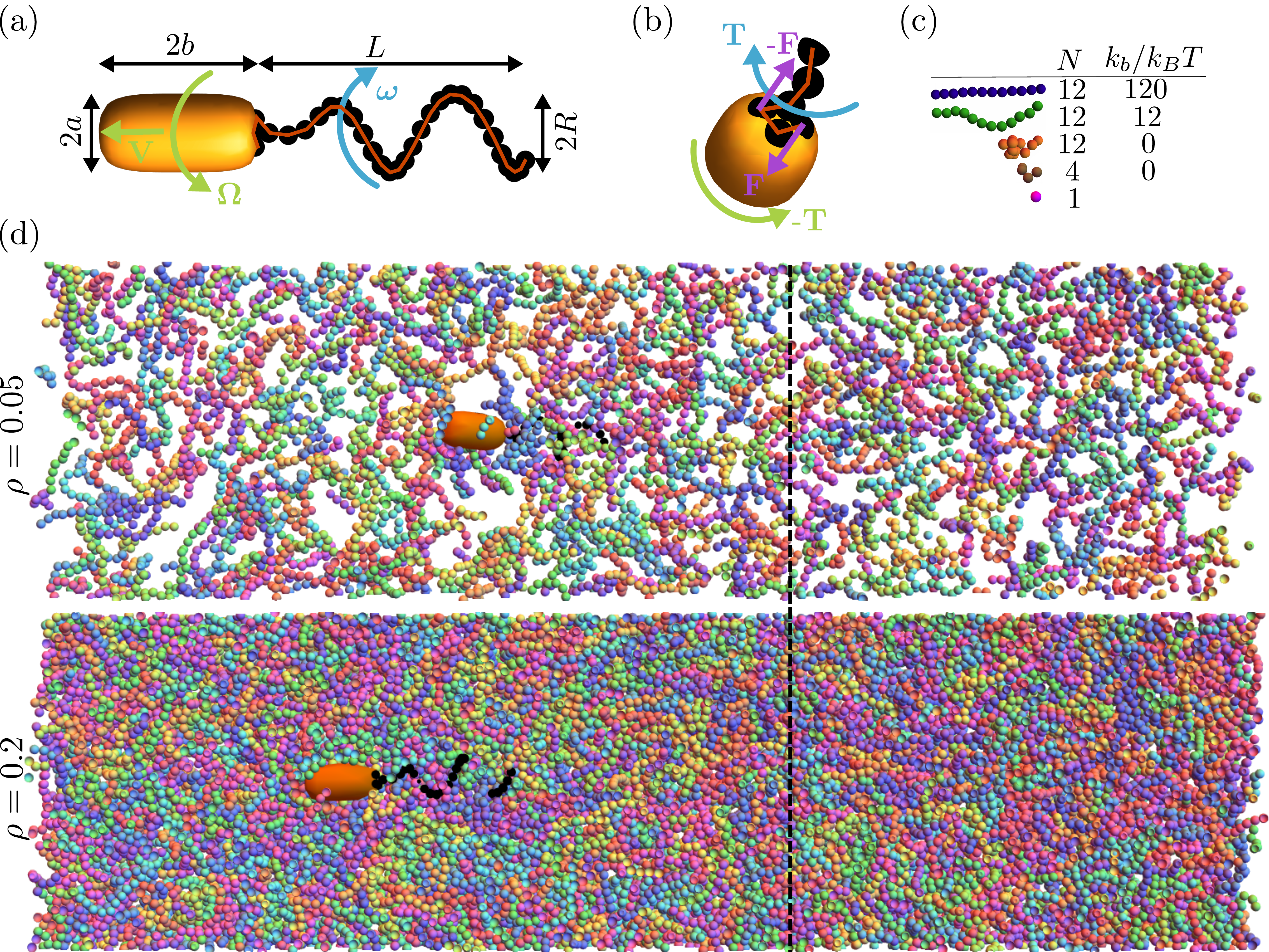}
\caption{\textbf{Simulations of a bacterium swimming in a dense macromolecular polymer solution.} (a) The bacterium consists of an elongated cell body with a right-handed helical flagellum attached.  (b) Body -- helix connection: a pair of forces  creates a torque $\mathbf{T}$ on the helix and an opposing torque  $-\mathbf{T}$ on the cell body. (c) Sketch of different polymers considered, of length $N$ and with bead-bead bending stiffness $k_b$ measured in units of the thermal energy $k_BT$ (Eq.~\ref{Eq:BendP}). (d) Typical simulation snapshots of the bacterial dynamics in fluids consisting of a Newtonian background fluid and including semiflexible polymers ($N=12$, $k_b=12k_BT$) at volume fraction $\rho=0.05$ (top) and $\rho=0.2$ (bottom).  The starting position of the cell body is indicated by the black dashed line showing that the bacterium swims faster in the more concentrated suspension. The colours of individual polymers are to aid visualisation. }
\label{Fig:1}
\end{center}
\end{figure}

\begin{figure}
\begin{center}
\includegraphics[width=.6\columnwidth]{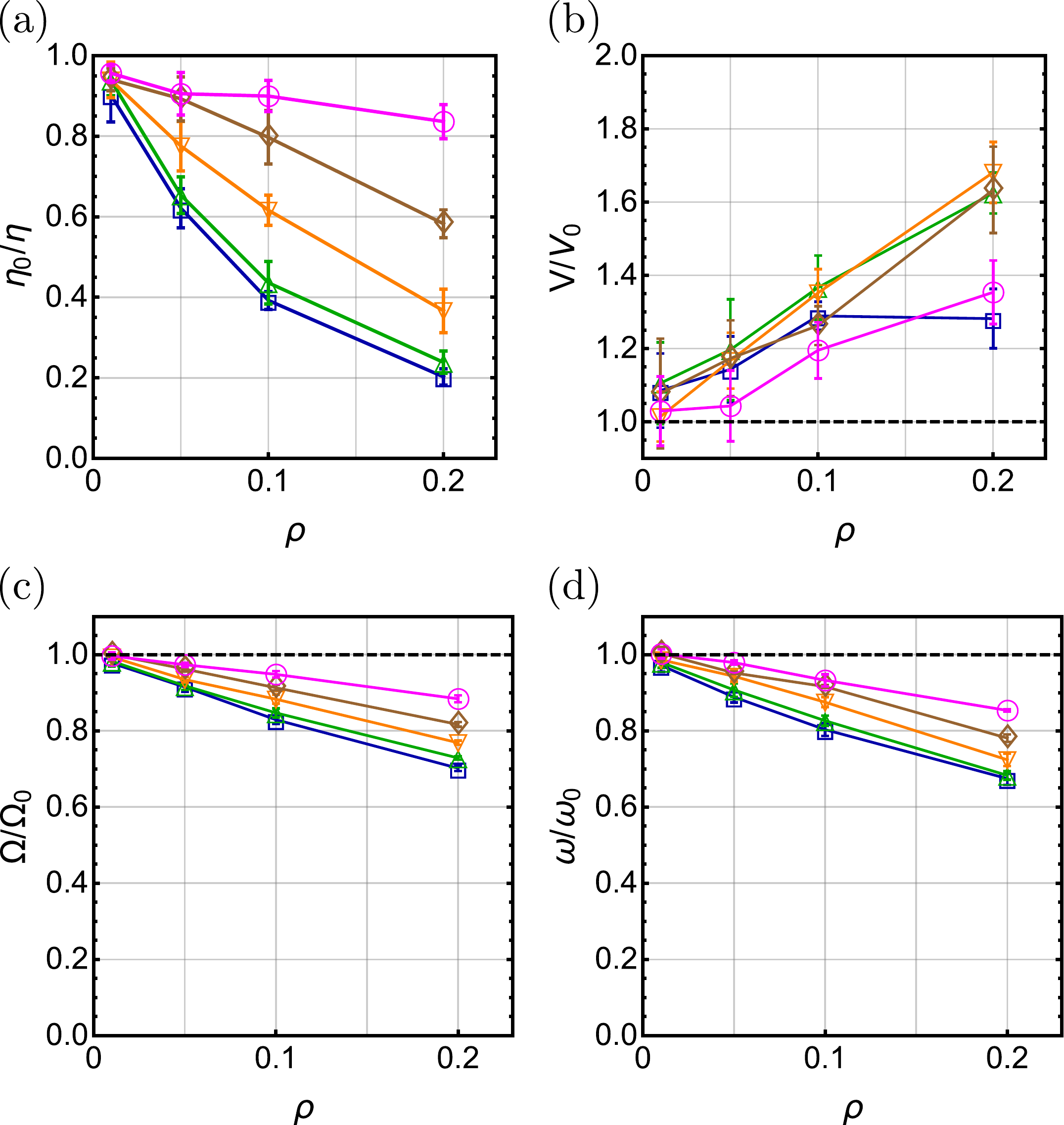}
\caption{\textbf{Fluid viscosity and swimming performance depend on polymer density}  (a) Fluidity (inverse viscosity) of polymer solutions. Error bars show the standard deviation from 45 time-averaged MPCD simulations.  (b) Average swimming speed. (c) Cell body angular velocity. (d) Flagellar angular velocity. Error bars in (b-d) are standard deviations from five time-averaged simulation runs. Colour code represents different polymer types and is the same as in Fig.~\ref{Fig:1}c. All quantities are scaled by their value in the polymer-free solution.}
\label{Fig:2}
\end{center}
\end{figure}

\begin{figure}
\begin{center}
\includegraphics[width=.8\columnwidth]{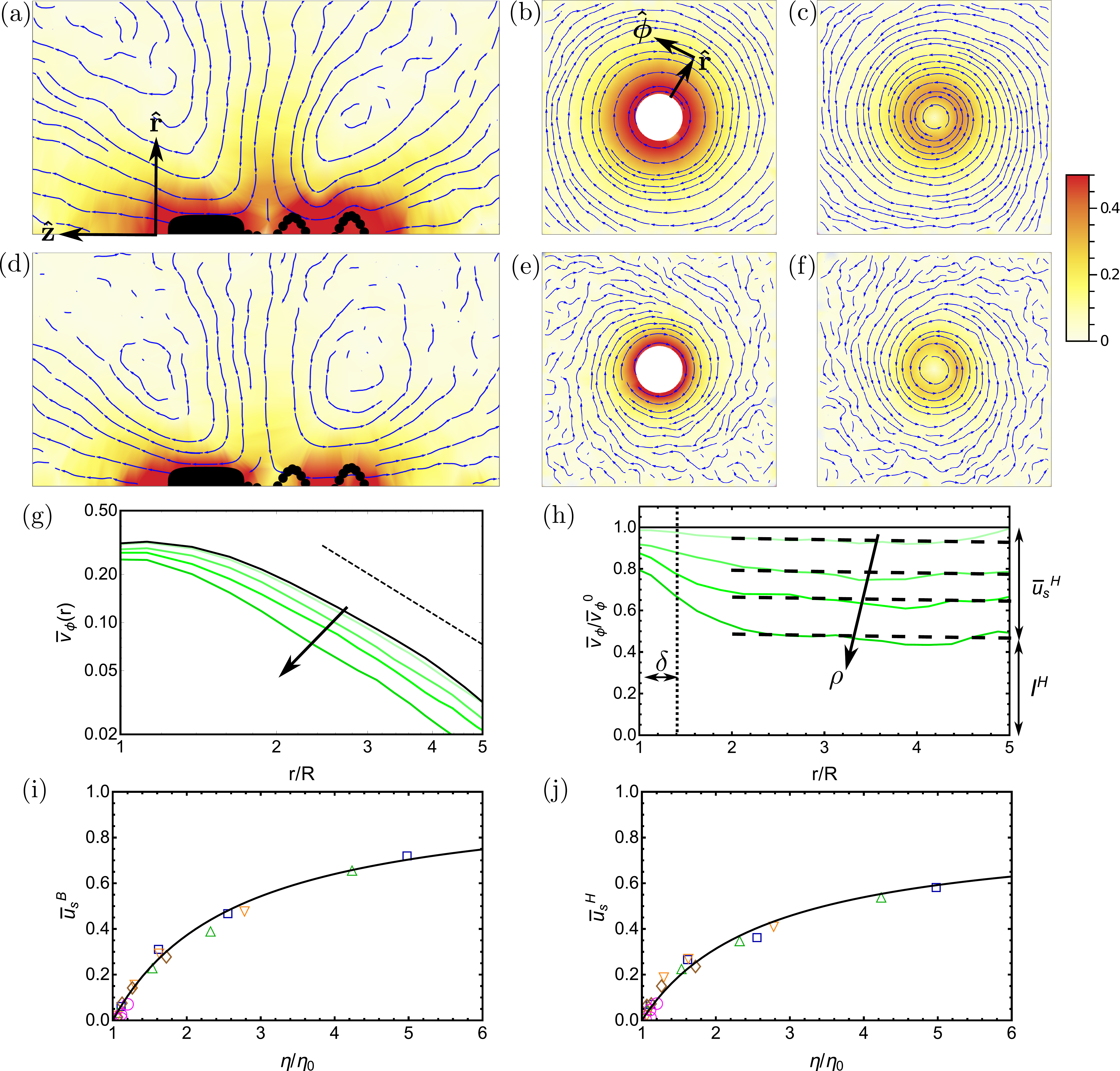}
\caption{\textbf{Flow fields and apparent slip velocities.} (a-f) Time- and ensemble averaged flow fields around the bacterium in the absence  of polymers (a-c) and in the presence of dense semiflexible polymers (length $N=12$, bending stiffness $k_b=12k_BT$, volume fraction $\rho=0.2$) (d-f). (a,d) Azimuthally averaged flow fields in the $r$-$z$ plane normalized by  bacterium swimming speed $V$. (b,e) Flow fields around the center of the cell body projected onto the $\phi$-$r$ plane normalized by the body rotation velocity $a\Omega$. (c,f) Flows around the center of the flagellum projected onto the $\phi$-$r$ plane normalized by its rotation velocity $R\omega$. Stream lines in (a-f) are shown in blue, and the background colour  represents the strength of the flow fields (see scale bar). (g) Decay of the  azimuthal flow fields around the flagellum for different polymer densities $\rho=\{0.01,0.05,0.1,0.2\}$ normalized by $R\omega$. The dashed line shows the scaling $r^{-2}$. (h) Same flow field data as in (g) but shown divided by the flow fields in the polymer-free solvent case. These curves  level off to a constant, $I$, indicated by the black dashed lines  and define the scaled apparent slip velocities.  The dotted line indicates the estimated depletion layer thickness $\delta$. (i,j) Measured scaled apparent slip velocities near the cell body ($\bar{u}_s^B=u_s^B/(a\Omega)$,(i)) and flagellum ($\bar{u}_s^H=u_s^H/(R\omega)$,(j)) for different polymeric fluids plotted versus scaled viscosity. Colour  code as in Fig.~\ref{Fig:1}c. The black lines show the results obtained from the theoretical model (see Methods).
}
\label{Fig:3}
\end{center}
\end{figure}

\begin{figure}
\begin{center}
\includegraphics[width=.65\columnwidth]{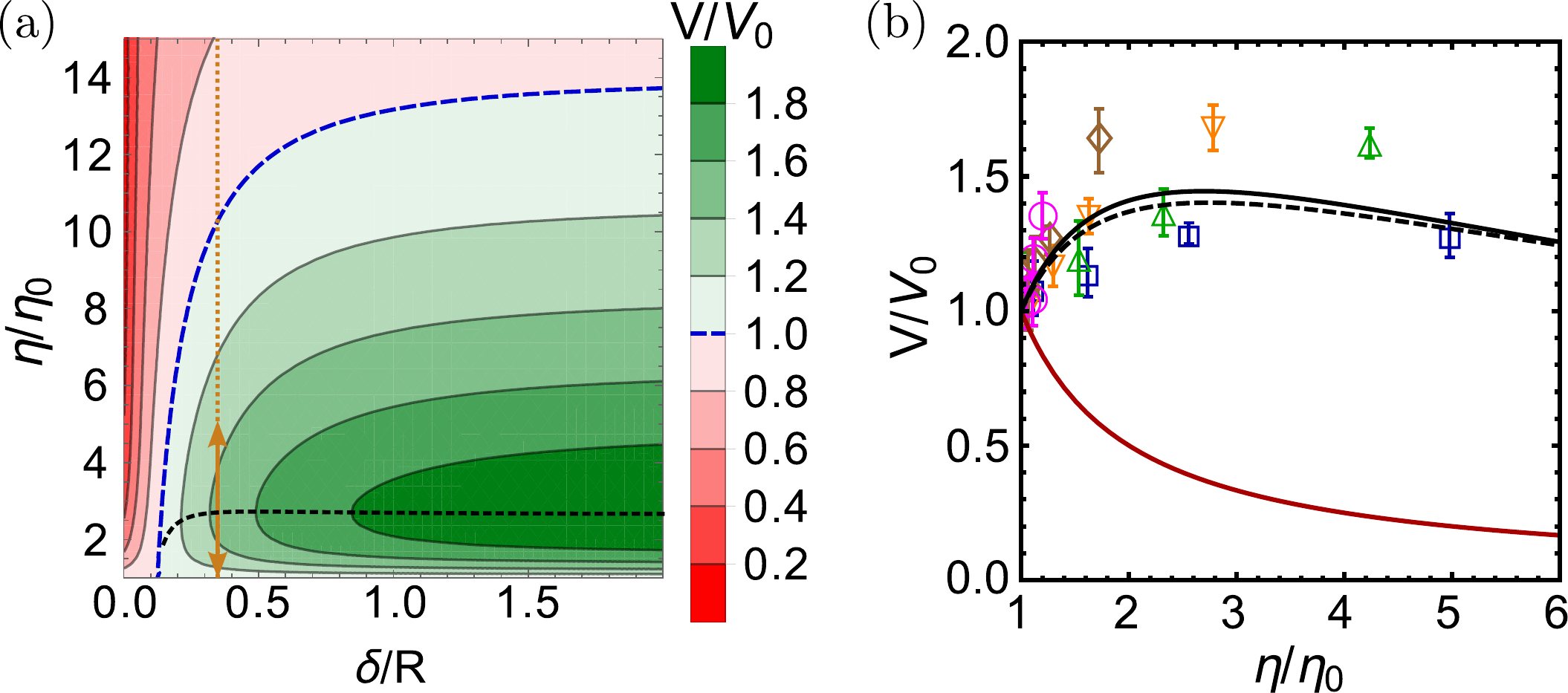}
\caption{\textbf{Dependence of swimming speed on fluid properties.}
(a) Dependence of  swimming speed obtained from analytical model [see Methods, Eq.~(\ref{Eq:FullM3})] on  fluid viscosity  and   thickness of the depletion layer scaled by the flagellum radius around the swimmer. The dotted orange line shows the measured depletion layer thickness $\delta=0.35R$ and arrows indicate the viscosity range obtained from MPCD simulations. The dashed blue line shows the transition between speed enhancement and speed reduction, and the dotted black line shows the viscosity corresponding to maximum speed enhancement for a given $\delta$.  (b) Comparison between model and simulations: The symbols show simulation results for swimming in different fluids (color  code as in Fig.~\ref{Fig:1}c).  The error bars are the same as shown in Fig.~\ref{Fig:2}b. The black dashed line shows the theoretical curve for a torque-driven helix including apparent slip [Eq.~(\ref{Eq:v0})], and the black solid line shows the change when the body is included [see Methods, Eq.~(\ref{Eq:FullM3})].  The red line indicates the theoretical prediction for a viscous continuum fluid without a polymer depleted region. Velocities and viscosities are scaled by their values in the polymer-free solution. }
\label{Fig:4}
\end{center}
\end{figure}

\renewcommand{\thefigure}{S\arabic{figure}}
\setcounter{figure}{0}

\begin{figure}[hbt]
\begin{center}
\includegraphics[width=.9\columnwidth]{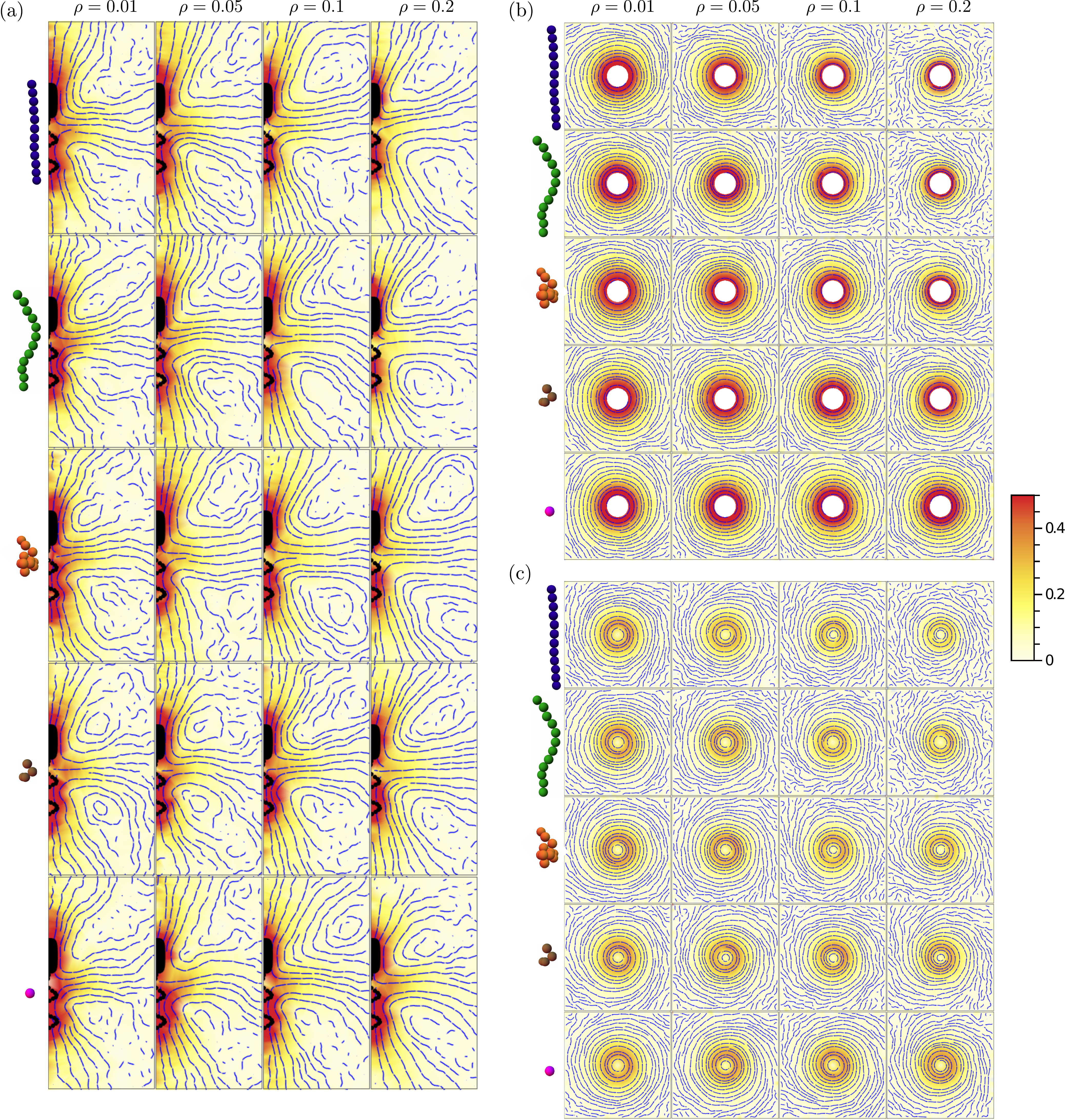}
\caption{\textbf{Flow fields around the bacterium for different polymeric fluids.}
Time- and ensemble averaged flow fields around the bacterium swimming in different polymeric fluids. Sketches of polymer types are the same as in Fig.~1c of the main text. (a) Azimuthally averaged flow fields in the $r$-$z$ plane normalized by  swimming speed $V$. (b) Flow fields around the centre of the cell body projected onto $\phi$-$z$ plane normalized by body rotation velocity $a\Omega$. (c) Flows around the centre of the flagellum projected onto $\phi$-$z$ plane normalized by its rotation velocity $R\omega$. Stream lines  are shown in blue, and the background colour represents the strength of the flow fields (see scale bar). }
\label{Fig:S1}
\end{center}
\end{figure}

\begin{figure}[hbt]
\begin{center}
\includegraphics[width=0.95\columnwidth]{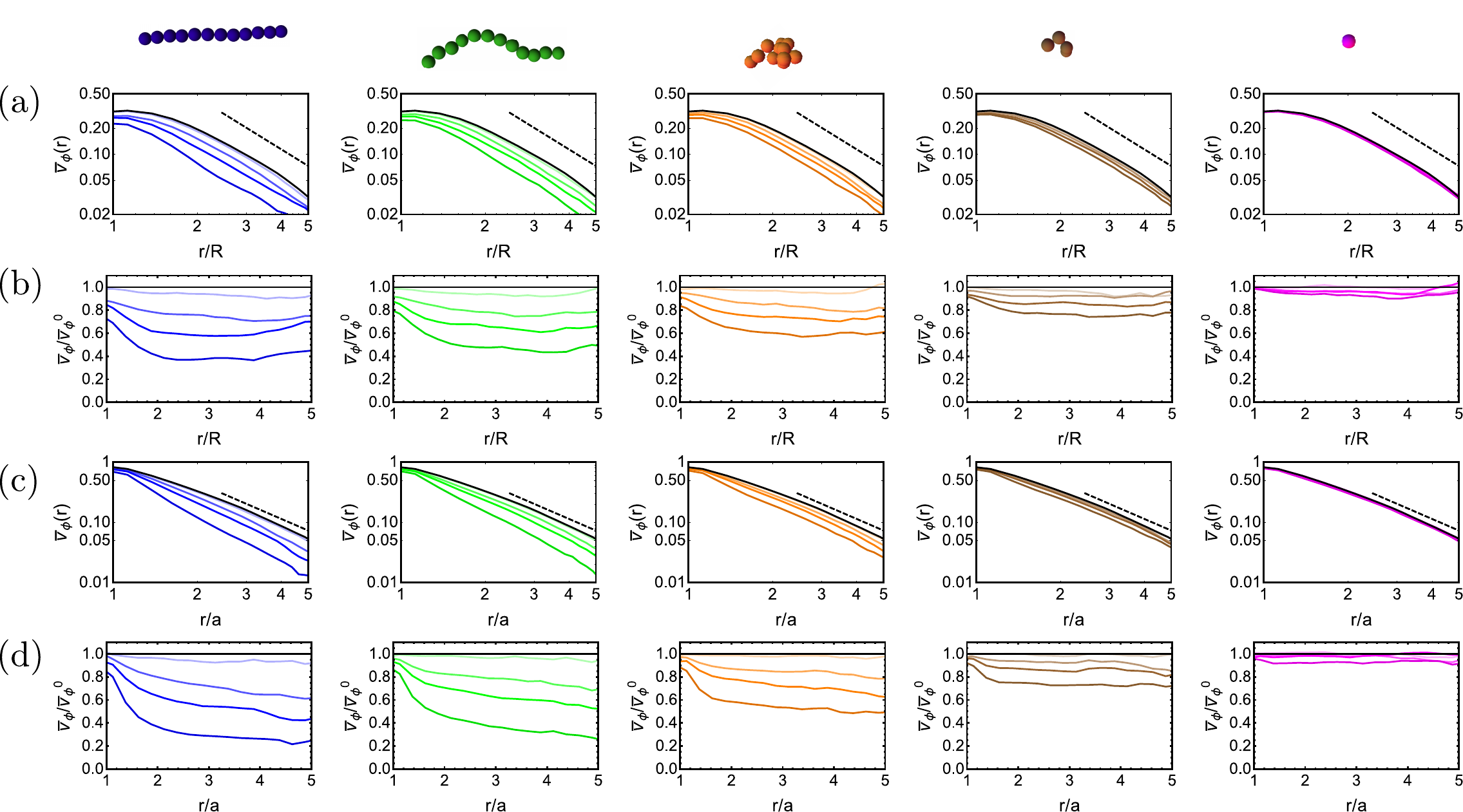}
\caption{\textbf{Deacay of azimuthal flow fields for different polymeric fluids.}
(a,c) Decay of the scaled azimuthal flow fields around the cell body (a) and the flagellum (c) for different polymer densities $\rho=\{0.01,0.05,0.1,0.2\}$. The dashed line shows the scaling $r^{-2}$. (b,d) Same flow field data as in (a,c) but shown divided by the flow fields in the polymer-free solvent case. }
\label{Fig:S2}
\end{center}
\end{figure}

\begin{figure}[hbt]
\begin{center}
\includegraphics[width=.65\columnwidth]{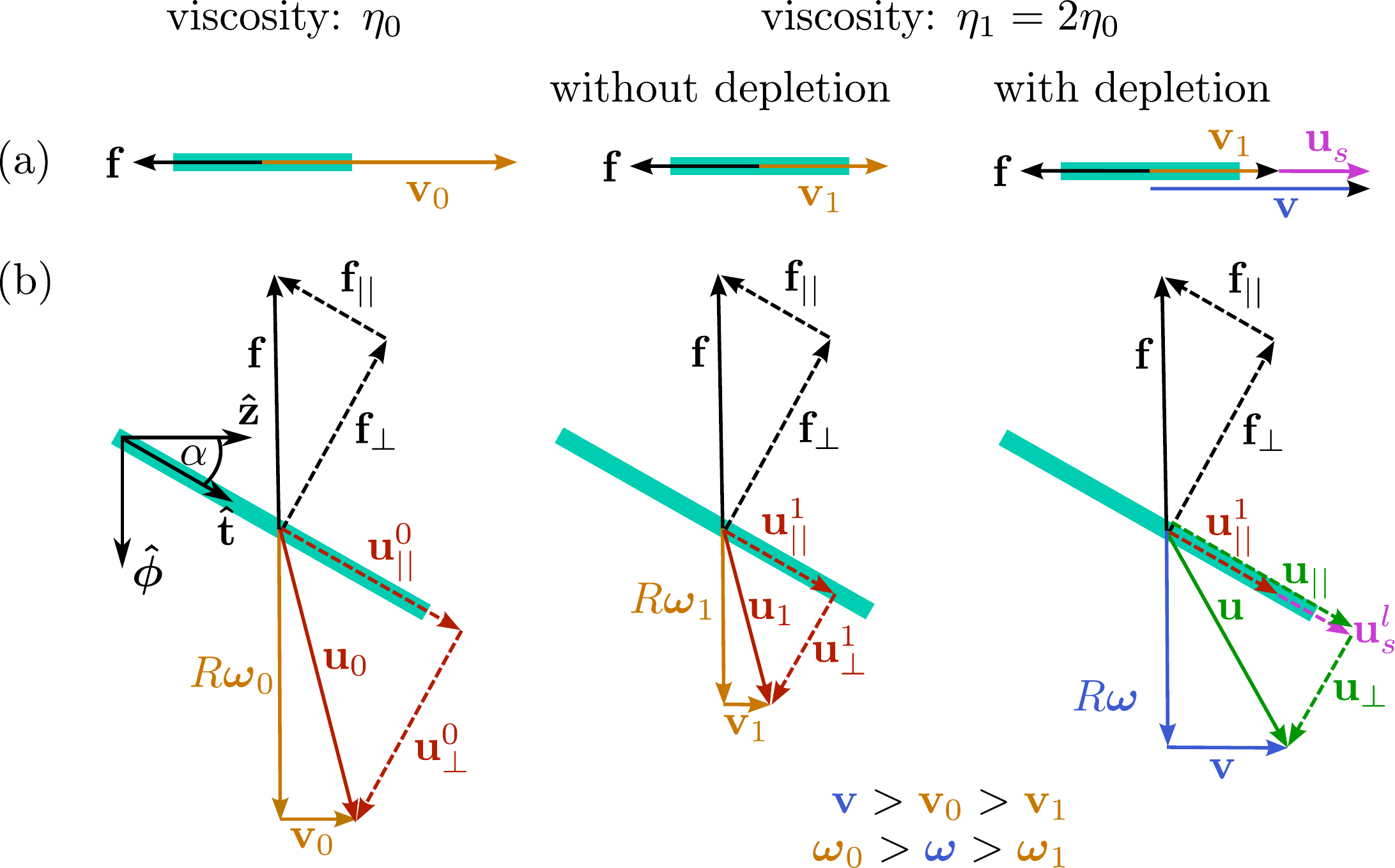}
\caption{\textbf{Resistive Force Theory (RFT) for driven rod and torque-driven helix}:
RFT for a rod driven along its axis (a) and for a torque-driven helix (b) in a  fluid of viscosity $\eta_0$ (left) and of  viscosity $\eta = 2\eta_0$ without (middle) and with (right) including polymer depletion represented by an apparent slip velocity. The respective parallel and perpendicular friction coefficients of a rod per unit length are denoted by $\xi_{||}^{0}$ and $\xi_{\perp}^{0}$ at viscosity $\eta_0$ and by $\xi_{||}^{1}=(\eta/ \eta_0)\xi_{||}^{0}$ and $\xi_{\perp}^{1}=(\eta/ \eta_0)\xi_{\perp}^{0}$ at viscosity $\eta_1$. The constant viscous force per unit length  opposing the motion of the rod is denoted by $\mathbf{f}$.  (a)  RFT  for the motion in a continuum viscous fluid with viscosity $\eta_0$, $\mathbf{f} = -\xi_{||}^{0}\mathbf{v}_0$ (left) and with viscosity $\eta_1$, $\mathbf{f} = -\xi_{||}^{1}\mathbf{v}_1$ (middle); right: motion in a polymer solution of bulk viscosity $\eta$ including an apparent slip, $\mathbf{f} = -\xi_{||}^{1}(\mathbf{v} - \mathbf{u}_s) = -\xi_{||}^{1}\mathbf{v}_1$. Here the rod velocity is still smaller than without polymers, $|\mathbf{v}| < |\mathbf{v}_0| $. (b) RFT for torque-driven helix of radius $R$. Torque is modeled by a uniform force density $\mathbf{f}$ acting along the $\boldsymbol{\hat{\phi}}$ direction of the helix; motion in a continuum viscous fluid with viscosity $\eta_0$, $\mathbf{f} = -\xi_{||}^{0}\mathbf{u}_{||}^0 - \xi_{\perp}^{0}\mathbf{u}_{\perp}^0$ leading to angular velocity $\boldsymbol{\omega}_0$ and velocity $\mathbf{v}_0$ (left), and with viscosity $\eta_1$, $\mathbf{f} = -\xi_{||}^{1}\mathbf{u}_{||}^1 - \xi_{\perp}^{1}\mathbf{u}_{\perp}^1$ leading to angular velocity $\boldsymbol{\omega}_1$ and velocity $\mathbf{v}_1$ (middle); right: motion in a polymer solution of bulk viscosity $\eta$ including apparent slip: $\mathbf{f} = -\xi_{||}^{1}(\mathbf{u}_{||} - \mathbf{u}_s^l) - \xi_{\perp}^{1}\mathbf{u}_{\perp}$ leading to angular velocity $\boldsymbol{\omega}$ and velocity $\mathbf{v}$. Although the angular velocity is reduced in the presence of polymers ($|\boldsymbol{\omega}| < |\boldsymbol{\omega}_0|$), the velocity of the helix increases: $|\mathbf{v}|>|\mathbf{v}_0|$. }
\label{Fig:S3}
\end{center}
\end{figure}

\begin{figure}[hbt]
\begin{center}
\includegraphics[width=.9\columnwidth]{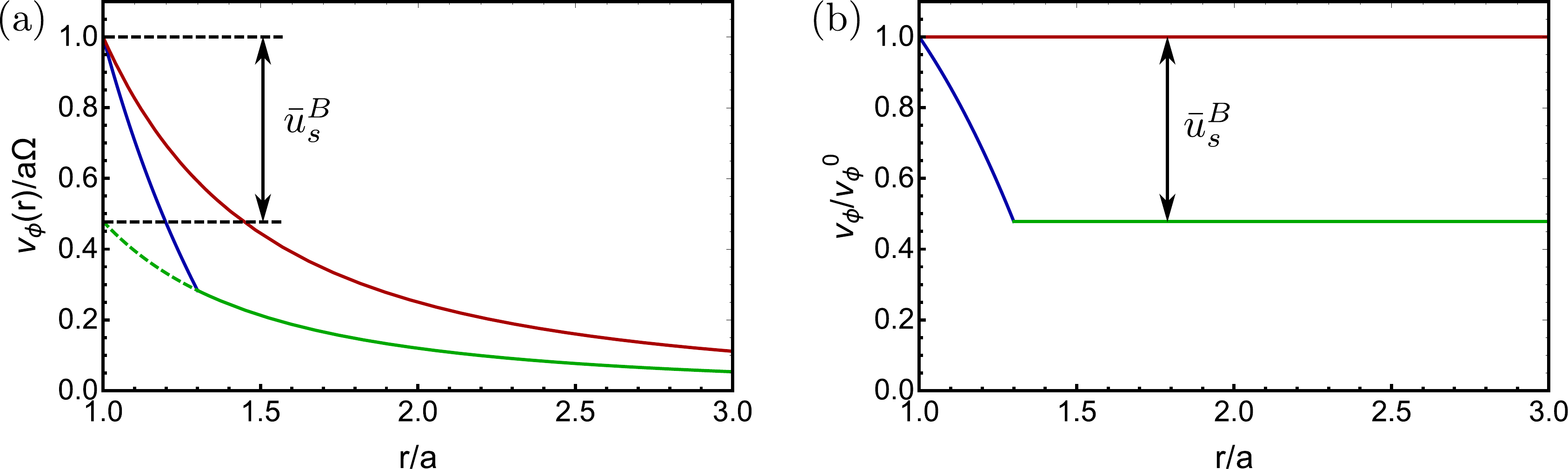}
\caption{\textbf{Flow field around rotating sphere in two-fluid model}: 
(a) Scaled flow field in a continuum viscous fluid at viscosity $\eta_0$ (red curve), and in a fluid consisting of two layers of viscosity $\eta_0$ at distances $r < a+\delta$ (blue curve)  and of viscosity $\eta=3\eta_0$ at $r > a+\delta$ (green curve) with $\delta=0.3a$. (b) Flow field relative to flow field in the polymer-free solvent. It decays for $r < a+\delta$ (blue curve) and is constant for $r > a+\delta$ (green curve) which defines the apparent slip velocity. The red curve shows the limit $\eta \rightarrow \eta_0$ or $\delta \rightarrow 0$ where the apparent slip vanishes. }
\label{Fig:S4}
\end{center}
\end{figure}

\begin{figure}[hbt]
\begin{center}
\includegraphics[width=.45\columnwidth]{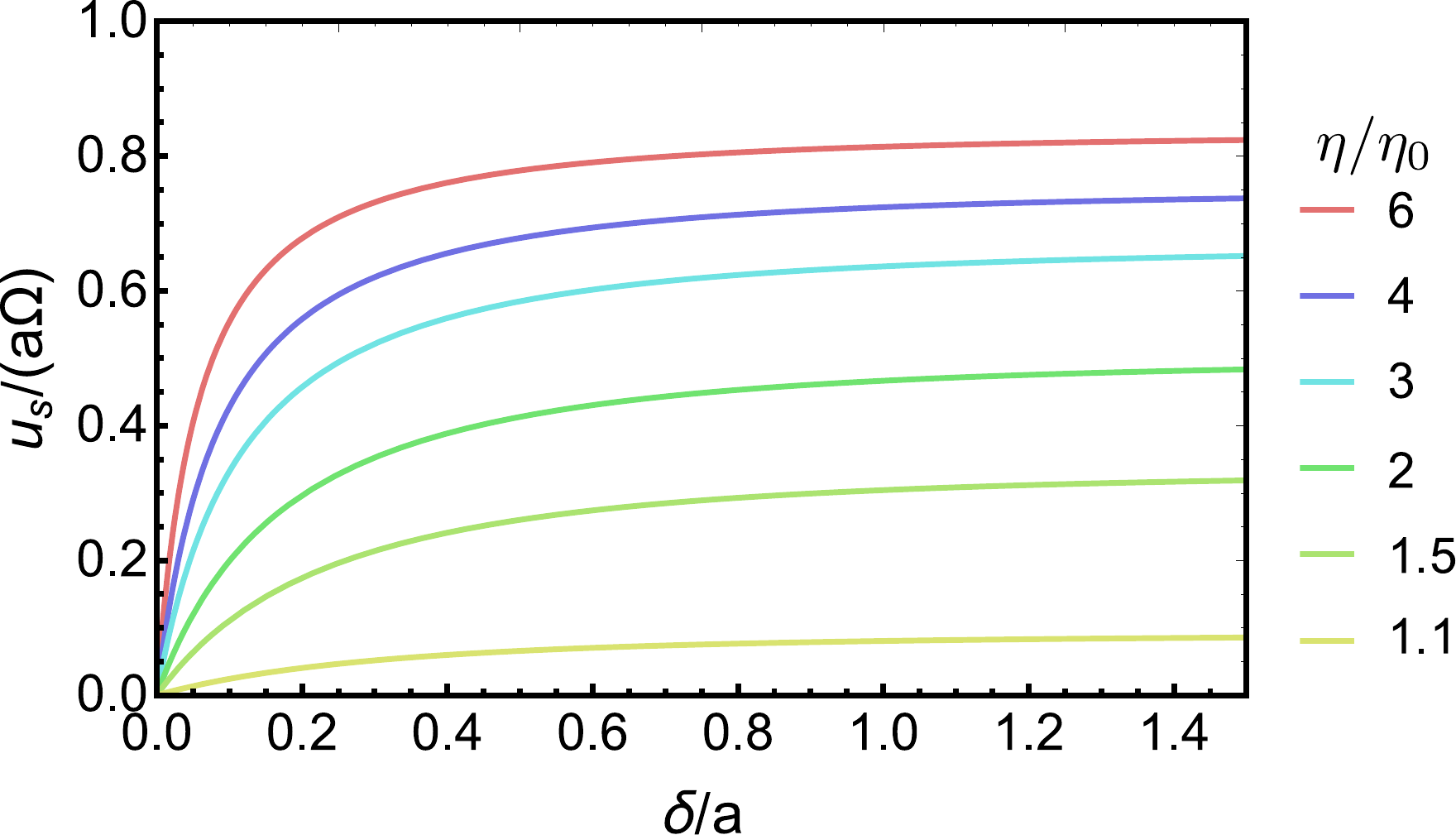}
\caption{\textbf{Apparent slip velocity model -- dependence on parameters}: How the scaled apparent slip velocity $u_s/(a\Omega)$ at the surface of a rotating sphere of radius $a$ rotating with angular velocity $\Omega$  depends on  depletion layer thickness $\delta$ for different viscosities $\eta$ (see Methods, Eq.~(17)). }
\label{Fig:S5}
\end{center}
\end{figure}


\end{document}